\documentclass[12pt]{article}
\usepackage{amsmath}
\begin{document}
\title{The electrostatic  potential of a periodic lattice}
\author{ G.Vaman  \thanks{vaman@ifin.nipne.ro} \\Institute of Atomic Physics, P.O.Box MG-6, Bucharest, 
Romania}
\maketitle
\abstract
We calculate the electrostatic
potential of a periodic lattice of arbitrary extended  charges
 by using the Cartesian multipole formalism. This method allows the 
 separation of the long-range potential from the contact potential 
 (potential on the source). We write both the electrostatic potential and the interaction energy as convergent sums in the reciprocal space.
\endabstract

\section{Introduction}
The calculation of the electrostatic potential of a periodic lattice is an
essential step in the study of crystals. For finding the electron energy bands,
the Schrodinger and Poisson equations must be solved self-consistently. This
subject has been extensively studied in the literature, by different methods
 (see Ref.\cite{kho} and references therein).
In this
paper we are concerned with the problem of finding the electrostatic potential
when the periodic charge distribution is known 
and we  are going to use the  multipolar formalism, as presented in
Refs. \cite{dub}, \cite{rad}, in order to solve this problem.
Unlike the Fourier technique, our method has the advantage of separating the
quantities which are
responsible for the contact interaction and which give us information about the
shape and spatial extension of the sources (mean radii)
from those which are responsible for the long-range
interaction (charge moments). Thus, we are able to correctly separate the
overlapping electrostatic energy from the long-range electrostatic energy and to
write both of them as convergent series in the reciprocal space. 

Our method can be used 
either together with an ab-initio model of the charge density  or in the formalism of pseudo-atom model.
We suppose bellow that we have decided on the form of the charge density
attached to a simple Bravais lattice and present the method of calculation of
the electrostatic potential and energy. We are considering a simple Bravais
lattice for simplifying the formulas, but the formalism can be easily extended
to an arbitrary periodic lattice.

The paper is organized as follows: in the next section we present the general
mathematical formalism which can be used for a Bravais  lattice with
extended arbitrary charges. In Sec.3 we study some particular simple charge
distributions, with a particular pedagogical emphasis on the lattice of
uniformly charged spheres and the end section is devoted to conclusions.

\section{General mathematical formalism}

We have proved in Ref.\cite{rad} the following formula for the Cartesian
multipole expansion of an arbitrary charge distribution:
\begin{align} \label{ch}
\rho({\bf r},t)=  \sum_{l=0}^{\infty} \sum_{n=0}^{\infty} \frac{(-1)^l (2l+1)!!}{
2^n n! l! (2n+2l+1)!!} \cdot \nonumber \\ \overline{r_{i_1}^{2n}}_{\dots i_l}(t) \Delta^n
\partial_{i_1} \dots \partial_{i_l} \delta( {\bf r}),
\end{align}
where $\rho({\bf r},t) $ is the charge density and 
\begin{equation} \label{rad}
\overline{r_{i_1}^{2n}}_{\dots i_l}(t)  = \frac{(-1)^l}{(2l-1)!!} \int d^3 {\bf \xi}
\xi^{2l+2n+1} \rho( {\bf \xi}, t) \partial_{i_1} \dots \partial_{i_l}
\frac{1}{\xi}
\end{equation}
is the electric mean square radius of order n and multipolarity l. The electric
mean square radii of order zero are also called  electric moments\footnote{ The
sum from the right-hand side of Eq. (\ref{ch}) begins with the monopole which
corresponds to $l=0$. As the monopole is a scalar, it doesen't need a dummy 
lower
index. So, Eq.(\ref{ch}) must be understood as follows:
$ \rho( {\bf r}, t) = \sum_{n=0}^{\infty} \frac{1}{2^n n! (2n+1)!} 
\overline{r_0^{2n}} \delta( {\bf r}) - \sum_{n=0}^{\infty}\frac{3}{2^n n!
(2n+3)!!}\overline{r_{i_1}^{2n}} \Delta^n \partial_{i_1} \delta( {\bf r})
 + \dots$} .
The multipoles in Eq. (\ref{ch}) are calculated with respect to the origin of
the coordinate axis.
>From the
above formulas, one can easily calculate the scalar potential by using
the formula:
\begin{equation}
\phi( {\bf r},t) = \int d {\bf r'} \frac{ \rho( {\bf r'},t)}{ |{\bf r} - {\bf r'}|}.
\end{equation}
After integating by parts one obtains:
\begin{align}\label{pote}
\phi( {\bf r},t)= \sum_{l=0}^{\infty} \sum_{n=0}^{\infty} \frac{(-1)^l (2l+1)!!}{
2^n n! l! (2n+2l+1)!!}\cdot \nonumber \\
\overline{r_{i_1}^{2n}}_{\dots i_l}(t) \Delta^n
\partial_{i_1} \dots \partial_{i_l} \frac{1}{r}
\end{align}

We consider now a Bravais lattice and some extended identical charge
distributions placed on the lattice points. For atomic crystals, these 
charge distributions are the neutral atoms and for ionic crystals they are the
ions which are neutralized by a uniform background. If the crystal is formed
by different species of atoms (ions), we can view it as a superposition of
simple Bravais lattices.
We denote by ${\bf R}_k$ a lattice vector. 
The charge density distribution of the lattice can be written as:
\begin{align} \label{chlat}
\rho({\bf r},t)= \sum_{ {\bf R}_k}
\sum_{l=0}^{\infty} \sum_{n=0}^{\infty} \frac{(-1)^l (2l+1)!!}{
2^n n! l! (2n+2l+1)!!} \cdot \nonumber \\
\overline{r_{i_1}^{2n}}_{\dots i_l}(t) \Delta^n
\partial_{i_1} \dots \partial_{i_l} \delta( {\bf r} - {\bf R}_k),
\end{align}
where we have  caculated the multipoles for each charge distribution with
respect  to the origin ${\bf R}_k$. As the charge distributions are identical,
their multipoles are identical and they do not need any lattice index.
The total scalar potential at an arbitrary point ${\bf r}$ of the
lattice is :
\begin{align}\label{potsf}
\phi({\bf r},t) =\sum_{ {\bf R}_k }
 \sum_{l=0}^{\infty} \sum_{n=0}^{\infty} \frac{ (-1)^l
(2l+1)!!}{ 2^n n! l! (2n+2l+1)!!}\cdot \nonumber \\
 \overline{r_{i_1}^{2n}}_{\dots i_l}(t) \Delta^n
\partial_{i_1} \dots \partial_{i_l} \frac{1}{ | {\bf r} - {\bf R}_k |}
\end{align}
We process further this expression by separating the contribution of the 
moments (long-range potentials) from the contribution of the mean-square
radii (contact potentials) and use the relations:
\begin{equation}\label{gen}
 \Delta \frac{1}{r} = - 4 \pi \delta( \vec{r} ),\; \; \; \sum_{\bf{R}_ k} \delta (
{\bf r} - {\bf R}_k ) = \frac{1}{\Omega} \sum_{ {\bf g} } e^{i {\bf g} {\bf r}},
 \end{equation}
where $\Omega$ is the volume of one elementary cell and ${\bf g}$ is a vector in
the reciprocal space. One obtains:
\begin{align}\label{sing}
\phi( {\bf r},t ) = & \sum_{ {\bf R}_k } \sum_{l=0}^{\infty} \frac{(-1)^l}{l!}
\overline{r_{i_1}^{0}}_{\dots i_l}(t) \partial_{i_1} \dots \partial_{i_l} \frac{1}{
| {\bf r} - {\bf R}_k | } - \nonumber \\
  & \frac{4 \pi}{ \Omega} \sum_{ {\bf g} } \sum_{l=0}^{\infty}
  \sum_{n=1}^{\infty} \frac{i^l (-1)^l (2l+1)!! (-1)^{n-1} g^{2n-2}}{ 2^n n! l!
  (2l+2n + 1)!!} \cdot \nonumber \\
  & \overline{r_{i_1}^{2n}}_{\dots i_l}(t) g_{i_1} \dots g_{i_l} e^{i
  {\bf g} {\bf r}}.
\end{align}

Now, using $W(t) = \int d {\bf r} \rho (
{\bf r},t) \phi({\bf r},t)$, after integrating by parts, we obtain the
 energy of the charge distribution placed at ${\bf r}={\bf R}_0$ in
the  potential $\phi( {\bf r},t ) $:
\begin{align*}
W(t)= & \sum_{l=0}^{\infty} \frac{1}{l!} \overline{r_{i_1}^{0}}_{\dots i_l}(t)
 \left( 
\partial_{i_1} \dots \partial_{i_l}(t) \phi({\bf r},t)  \right)_{ {\bf r} = {\bf
R}_0} \\
   & + \sum_{l=0}^{\infty} \sum_{n=1}^{\infty} \frac{(2l+1)!!}{ 2^n n! l!
   (2l+2n+1)!!}\cdot \\
   & \overline{r_{i_1}^{2n}}_{\dots i_l} (t)\left( \Delta^n
\partial_{i_1} \dots \partial_{i_l} \phi({\bf r},t)  \right)_{ {\bf r} = {\bf
R}_0}.
\end{align*}
The first term of r.h.s. the above formula is the long-range interaction energy,
 while
the second term  is different from zero only when $ \left(\Delta \phi(
{\bf r})\right)_{ {\bf r} = {\bf R}_0} \ne 0$, that is, when there exist non-zero
overlap between the charge at ${\bf r}= {\bf R}_0$ and some of the other
charges (contact or overlapping energy, including the self-energy).
The formulas of this section are valid for an arbitrary time-dependence, but in
the following we shall consider only the electrostatic case.

\section{Particular charge densities}
We apply the general formalism of the previous section for some particular charge
distributions. We shall analyse in greater detail the case of a uniformly
charged sphere and, for pedagogical purposes, we shall calculate the potential
of a uniformly charged sphere in two ways: first by using the spherical
multipole expansion of the Green function and then
 by using the  Cartesian multipole expansion of the charge density. Then, 
 we shall shortly present the
results for the potential and interaction energy for other interesting charge
distributions.

\subsection{Constant charge density}

\subsubsection{The potential of a uniformly charged sphere}
In many electrodynamics textbooks we can find the following method for
calculating the electrostatic potential of a uniformly charged sphere: from the
Poisson equation $ \Delta \phi ( {\bf r} ) = -4 \pi \rho ( {\bf r} )  $
one finds using the Green theorem:
\begin{equation}\label{green}
\phi ( {\bf r}) = \int d {\bf r'} \frac{ \rho( {\bf r'})}{ | {\bf r}- {\bf
r'}|}.
\end{equation} 
After inserting in the above equation the expansion:
\[
\frac{1}{ | {\bf r}- {\bf r'}|} = \sum_{l,m} \frac{4\pi}{2l+1}
\frac{r_<^l}{r_>^{l+1}} Y_{lm}(\theta, \phi) Y^*_{l,m}(\theta', \phi'),
\]
 using the fact that the charge density is spherically symmetric, one obtains:
\begin{equation}
\phi( {\bf r}) = 4\pi \int_0^{\infty} d r' r'^2 \rho (r') \Big[ \theta(r-r')
\frac{1}{r} + \theta(r'-r) \frac{1}{r'}  \Big].
\end{equation}
Using for the charge density the expression:
\begin{equation}\label{ch1}
\rho({\bf r}) = C  \theta(a-r),
\end{equation}
where $C$ is a constant and replacing $\theta(r-r') = 1- \theta(r'-r)  $,
after elementary integrations one finds:
\begin{equation}
\phi ( {\bf r}) = \frac{4 \pi a^3 C}{3r} +  4 \pi C ( \frac{a^2}{2} -
\frac{a^3}{3r} - \frac{r^2}{6})\theta(a-r).
\end{equation}
In order to separate the potential inside the sphere from the potential outside
it, one uses $1 = \theta(r-a) + \theta(a-r) $ and one finally finds the well
known result:
\begin{equation}\label{pof}
\phi( {\bf r}) = \frac{4 \pi a^3 C}{3r} \theta(r-a) + \frac{2\pi C}{3} (3a^2
-r^2) \theta(a-r).
\end{equation}

Now, we are going to obtain the same result for the electrostatic potential by
using the method of Cartesian multipole expansion. For the charge density 
(\ref{ch1}) one can easily see that  the only
non-zero multipoles are the charge monopole and its (infinite number of) mean
square radii.
After a simple calculation one finds the following expression of a charge mean
square radius of order $n$: 
\begin{equation} \label{rad1}
\overline{r_0^{2n}} = 4 \pi C \frac{a^{2n+3}}{2n+3}
\end{equation}
When we introduce Eqs. (\ref{ch1}) and (\ref{rad1}) in Eq. (\ref{ch}) we find:
\begin{equation}\label{piz1}
\theta(a-r) = 4\pi \sum_{n=0}^{\infty} \frac{a^{2n+3}}{2^n n! (2n+3)!!} \Delta^n
\delta({\bf r}).
\end{equation}
The above formula is a particular case of Eq. (\ref{ch}), which was rigorously
justified in Ref. \cite{rad}. On the other hand,
it can be formally obtained by integrating after "a" the
expansion: 
\begin{equation} \label{piz2}
\delta(a-r) = 4\pi \sum_{k=0}^{\infty} \frac{a^{2k+2}}{2^k k! (2k+1)!!} \Delta^k
\delta( {\bf r}), 
\end{equation}
which is a particular case of the mean value formula of Pizzetti (\cite{row}).
Note that both Eqs. (\ref{piz1}), (\ref{piz2}) can be justified in
the Fourier space. Thus, by taking the Fourier transform
$\int d {\bf r} e^{-i {\bf k}{\bf r}}  $ of Eq. (\ref{piz2})
one obtains:
\[
\frac{4\pi a}{k} sin(ka) = 4 \pi a \sum_{n=0}^{\infty} \frac{ (-1)^n a^{2n+1}
k^{2n+1}}{2^n n! (2n+1)!!},
\]
which, after using $ (2n+1)!! = \frac{ (2n+1)!}{2^n n!}  $, is exactly the
Taylor expansion of the function $sin(ka)$. If we take the Fourier transform of
Eq. ( \ref{piz1} ) we obtain:
\[
\frac{4\pi a^2}{k} j_1(ka) = 4\pi \sum_{n=0}^{\infty} \frac{ (-1)^n a^{2n+3}
k^{2n}}{2^n n! (2n+3)!!},
\]
which is exactly the Taylor expansion of the spherical Bessel function
$j_1(ka)$. By the same method of Fourier transforms one can justify the formula:
\begin{equation}\label{piz3}
\frac{\theta(a-r)}{r} = 4\pi \sum_{n=0}^{\infty} \frac{ a^{2n+2}}{
2^{n+1} (n+1)! (2n+1)!!} \Delta^n \delta( {\bf r}).
\end{equation}
The series of Dirac delta functions and its derivatives which appear in Eqs. (\ref{piz1}), 
(\ref{piz2}), (\ref{piz3}) are callel in the literature dual Taylor series and their correspondence with the usual Taylor series is rigorously studied in Ref. \cite{gel}.

We return now to our calculation of the electrostatic potential: inserting Eqs.
(\ref{ch1}), (\ref{piz1}) into Eq. (\ref{green}) one obtains after
integrating by parts:
\begin{equation}\label{po1}
\phi ( {\bf r} ) = \frac{4\pi a^3 C}{3r} - 16 \pi^2 C \sum_{k=1}^{\infty}
\frac{a^{2k+3}}{2^k k! (2k+3)!!} \Delta^{k-1} \delta( {\bf r} ).
\end{equation} 
The first term of the above equation is the contribution of the monopole to the
electrostatic potential and it is singular at $r=0$. The second term is the
contribution of the mean square radii to the electrostatic potential and  it
contains a singular contribution at $r=0$ (which exactly compensates the
contribution of the monopole) and a finite contribution which is exactly the
potential inside the sphere. Now, we use in Eq. (\ref{po1}) the relation $1=
\theta(r-a) + \theta (r+a)$ and Eq. (\ref{piz3}), in order to separate the
potential inside the sphere from the potential outside it. We obtain:
\begin {equation}\label{po2}
\phi({\bf r}) = \frac{4\pi a^3 C}{3r} \theta(r-a)
+ \frac{16 \pi^2 C}{3} \sum_{k=0}^{\infty}
\frac{a^{2k+5}}{2^k k! (2k+5)!!}\Delta^n \delta({\bf r}),
\end{equation}
which, after using (\cite{gel}): $ r^2 \Delta^n \delta( {\bf r} ) = 2n(2n+1)
\Delta^{n-1}  \delta( {\bf r} )  $, gives exactly Eq. (\ref{pof}).

As a final remark, note that all the contribution of the mean square radii is of
the form $f(r) \theta(a-r)$, that is, it is different from zero only inside the
sphere. One can see this writing Eq. (\ref{po1}) in the alternative form:
\begin{equation}
\phi({\bf r})= \frac{4 \pi a^3 C}{3r} + 4\pi C \Big( \frac{a^2}{2} -
\frac{a^3}{3r} - \frac{r^2}{6}  \Big) \theta(a-r)
\end{equation}
This observation is valid for any type of charge distribution and 
 is important when we calculate the overlapping interaction energy.

\subsubsection{Bravais lattice of uniformly charged spheres}
We consider a Bravais lattice and, at each lattice point,  a spherically charge
 density of the form Eq.(\ref{ch1}).
We do not impose any restriction on the sphere's radius $a$, so the spheres may
overlap each other. 
The electrostatic potential at an arbitrary point ${\bf r}$ of such a lattice
can be written:
\begin{align*}
\phi( {\bf r} )  = &\sum_{ {\bf R}_k} \frac{ \overline{r_0^0}}{|{\bf r} - {\bf
R}_k |} + \sum_{ {\bf R}_k} \sum_{n=1}^{\infty} \frac{ \overline{r_0^{2n}}}{ 2^n
n! (2n+1)!!} \Delta^n \frac{1}{|{\bf r} - {\bf R}_k |} = \\
              &  4 \pi C \frac{a^3}{3}
		\sum_{ {\bf R}_k} \frac{ 1}{|{\bf r} - {\bf R}_k |}  -  \\
		&  2 \pi C
		a^5 \sum_{ {\bf R}_k} \sum_{p=0}^{\infty} \frac{a^{2p}}{2^p
		(p+1)! (2p+5)!!} \Delta^p \delta ( {\bf r} - {\bf R}_k),
\end{align*}
where we have used the first relation of Eq.(\ref{gen}).  We can
further transform the above formula by using
Eq.(\ref{gen}). One obtains:
\begin{align}\label{pot1}
\phi( {\bf r}) =& \frac{4 \pi C a^3}{3} \sum_{ {\bf R}_k} \frac{1}{| {\bf r} - {\bf
R}_k|} + 2 \pi C a^5 \cdot \nonumber \\
&\sum_{ {\bf g}} \sum_{p=0}^{\infty} \frac{(-1)^p (ag)^{2p}}{2^p
(p+1)! (2p+5)!!} e^{i {\bf g} {\bf r}}= \nonumber\\
   & \frac{4 \pi C a^3}{3} \sum_{ {\bf R}_k} \frac{1}{| {\bf r} - {\bf
R}_k|}  + \frac{16 \pi^2 C a^3}{\Omega} \cdot \nonumber \\
& \sum_{ {\bf g}} ( \frac{1}{ag^3} j_1(ag) - \frac{1}{3g^2}) 
 e^{i {\bf g} {\bf r}}.
\end{align}
At this stage, we are going to make some general remarks about the structure of
the electrostatic potential, as it appears in the above equation. 
The first term in the r.h.s. of the above equation is the monopole contribution to the scalar
potential and the second term in the r.h.s. of Eq.(\ref{pot1})
is the contribution from all the monopole mean square radii. 
It is worth noting that, if we consider a Bravais lattice of neutral spheres
(which means a charge density of the form: $\rho({\bf r})= C \theta(a-r) -
\frac{4\pi a^3 C}{3} \delta( {\bf r})$), the first term of the above equation
dissapears. The second term is the potential inside the spheres and it is a superposition of potentials of the form
$f(r-{\bf R}_i)\theta(a-|{\bf r}- {\bf R}_i|)$. The term which corresponds to
${\bf g}=0$ is the so-called Bethe mean potential (\cite{bet}). 
We can see that, in fact,  the
mean electrostatic potential of the lattice 
 comprises the contribution of all the infinite number of
mean square radii. Only this complete potential, by Fourier transforming, 
will give us
correct information about the real shape of the charge distribution.
We can see this if we pass again to the real space
 taking the Fourier transform and
 applying the Poisson summation formula.
  One obtains:
\begin{align}\label{finr}
\phi({\bf r})=& \sum_{ {\bf R}_k } \Bigg[
\frac{4\pi C a^3}{3| {\bf r} - {\bf R}_k|} \theta(  | {\bf r} - {\bf R}_k| -a) + \nonumber \\
&\frac{2\pi C}{3} (3a^2 -| {\bf r} - {\bf R}_k|^2)
 \theta( a - | {\bf r} - {\bf R}_k| ) \Bigg],
\end{align}
which is exactly the superposition of the potentials of all the spheres from the
lattice points. This is the expected result, as the electrostatic potential
satisfies the superposition principle.

It is well-known that the series $\sum_{ {\bf R}_k} \frac{1}{| {\bf r} - {\bf
R}_k|}$ is conditionally convergent and there exist many methods in the
literature for studying it (see Refs.  \cite{kho}, \cite{har} and references
therein). As we are not very much interested here about what happens at
infinity (surface effects), we shall consider this term as arising from the limit of a screened
Coulomb potential, whose Fourier transform is well defined at the origin.
Therefore, we shall write Eq.(\ref{pot1}) in the reciprocal space as follows:
\begin{align}\label{con}
\phi( {\bf r}) =&
    \frac{16 \pi^2 C a^3}{3} \lim_{\epsilon \rightarrow 0} \sum_{ {\bf g}}
    \frac{e^{i {\bf g} {\bf r}}}{g^2+\epsilon^2}
     + \frac{16 \pi^2 C a^3}{\Omega} \cdot \nonumber \\
& \sum_{ {\bf g}} ( \frac{1}{ag^3} j_1(ag) - \frac{1}{3g^2}) 
 e^{i {\bf g} {\bf r}}.
\end{align}
As the self-energy of a uniformly charged sphere is 
$W_s= \frac{16 \pi^2C^2 a^5}{15},$
 then the electrostatic energy of the lattice is:
\begin{align} \label{self}
W= & \frac{32 \pi^3 C^2 a^5}{3 \Omega} 
 \lim_{\epsilon \rightarrow 0}\sum_{ {\bf g}} \frac{j_1
(ag)}{g(g^2+\epsilon^2)} + \frac{32 \pi^3 C^2 a^5}{\Omega} \cdot \nonumber \\
& \sum_{{\bf g}} \left[  
\frac{(j_1(ag))^2}{ag^4} - \frac{j_1(ag)}{3g^3} \right]
- \frac{16 \pi^2C^2 a^5}{15}
    \end{align}
and, after twe take the limit:
\begin{equation}
W=\frac{32 \pi^3 C^2 a^4}{\Omega} \sum_{ {\bf g}} \Big[ \frac{ j_1(ag)}{g^2}
  \Big]^2 - \frac{16 \pi^2C^2 a^5}{15}.
\end{equation}

This result, which in our view is valid for arbitrary (overlapping or
non-overlapping spheres), coincide with the result from Ref. \cite{ber1}, 
Eq.(41). (In our paper the structure factor $ F({\bf g})=1$, because we
are considering a simple Bravais lattice). However, in Ref. \cite{ber1}
this result is reported as being valid only for nonoverlapping spheres
and is subsequently corrected for overlap by adding a sum in the direct
space. 
This discrepancy can be explained as follows:
the first term in the r.h.s. of  Eq.(\ref{self}) is the long-range interaction energy, while
the second is the overlapping energy. The second term is
therefore zero in the case of nonoverlapping spheres. It follows that, for
zero overlap we have:
\begin{equation}
\sum_{ {\bf g}} \frac{ (j_1(ag))^2}{ag^4} = \sum_{\bf g}
\frac{j_1(ag)}{3g^3}
\end{equation}
and the interaction energy can be written in two equivalent forms:
\begin{align*}
W = & \frac{32 \pi^3 C^2 a^5}{3 \Omega} \sum_{{\bf g}} \frac{
j_1(ag)}{g^3}- \frac{16 \pi^2C^2 a^5}{15} = \\
&\frac{32 \pi^3 C^2 a^4}{ \Omega} \sum_{{\bf g}}\frac{
(j_1(ag))^2}{g^4}- \frac{16 \pi^2C^2 a^5}{15}.
\end{align*}
Evidently, the model of overlapping spheres having uniform charge densities is counterintuitive, but our method of calculation can be applied for arbitrary charge densities, as we shall do in the next sections.

It is worth analyzing in more detail our Eq. (\ref{self}) and comparing our method with that of Ref. \cite{ber1}.
The method of charge spreading introduced in Ref.\cite{ber1} is based on a theorem of electrostatics which asserts that the electrostatic energy of two point charges coincide with the electrostatic energy of two spherically symmetric extended charges. If we look at our Eq.(\ref{self}), we see that this means to admit that the term $\lim_{\epsilon \rightarrow 0}\sum_{ {\bf g}} \frac{j_1 (ag)}{g(g^2+\epsilon^2)}$ exactly cancel the term $\sum_{{\bf g}}  \frac{j_1(ag)}{3g^3} $. This does not mean that we have solved on this occasion the problem of summing the conditionally Coulomb series but that using such a mathematical machinery we get widely accepted physical results. Having in mind this conclusion, we shall apply the same formalism in the next sections, for other interesting charge densities.

\subsection{Other charge distributions}

\subsubsection{Gaussian charge distribution}
We consider now that the charge density
has a gaussian form :
$\rho( {\bf r} ) = C e^{- \lambda r^2} $, where $C$ and $\lambda$ are adjustable
parameters. Then, the charge mean square radii of order $n$ has the form:
\begin{equation}
\overline{r_0^{2n}}= \frac{2\pi C}{\lambda^{n+\frac{3}{2}}} \Gamma(n+
\frac{3}{2}),
\end{equation}
where $\Gamma$ is the Gamma function (\cite{gra}).
>From Eq. (\ref{ch}) we obtain in this case:
\begin{equation}
e^{-\lambda r^2} = \pi^{3/2} \sum_{n=0}^{\infty} \frac{1}{ 2^{2n} n!
\lambda^{n+\frac{3}{2}}} \Delta^n \delta({\bf r}),
\end{equation}
which, after Fourier transformation becomes the Taylor expansion of the
exponential function:
\begin{equation}
e^{-\frac{k^2}{4\lambda}} = \sum_{n=0}^{\infty} \frac{1}{n!} \left(- \frac{k^2}{4
\lambda} \right)^n.
\end{equation}
 Then, for the electrostatic potential
produced by all the ions at the point ${\bf r}$ one obtains:
\begin{align}
\phi({\bf r}) = & \frac{\pi^{3/2} C}{\lambda^{3/2}} \sum_{ {\bf R}_k} \frac{1}{ |
{\bf r} - {\bf R}_k |} + \frac{4 \pi^{5/2} C}{\Omega \lambda^{3/2}} \cdot \nonumber \\
&\sum_{ {\bf
g}} \frac{1}{g^2} ( e^{- \frac{g^2}{4 \lambda}} -1) e^{i {\bf g} {\bf r}},
\end{align}
or, as a single sum in the reciprocal space:
\begin{align}
\phi({\bf r}) = & \frac{4 \pi^{5/2} C}{\Omega \lambda^{3/2}}
\lim_{\epsilon \rightarrow 0} \sum_{ {\bf
g}} \frac{ e^{i {\bf g} {\bf r}}}{g^2+\epsilon^2}    + 
\frac{4 \pi^{5/2} C}{\Omega \lambda^{3/2}} \cdot \nonumber \\
&\sum_{ {\bf g}} \frac{1}{g^2} ( e^{- \frac{g^2}{4 \lambda}} -1) e^{i {\bf g} {\bf r}}.
\end{align}
Taking the Fourier transform and using the Poisson summation formula, we obtain
from the above equation:
\begin{align*}
\phi({\bf r})= &\frac{ \pi^{3/2}C}{ \Omega \lambda^{3/2}} \sum_{{\bf R}_k}
\frac{ 1}{ | {\bf r} - {\bf R}_k|} + 
\frac{ \pi^{3/2}C}{ \Omega \lambda^{3/2}}  \cdot \\
&\sum_{{\bf R}_k} \left( \frac{erf(
\sqrt{\lambda} | {\bf r} - {\bf R}_k|)}{| {\bf r} - {\bf R}_k|} -    
\frac{ 1}{ | {\bf r} - {\bf R}_k|}  \right)\\
&= \frac{ \pi^{3/2}C}{ \Omega \lambda^{3/2}}  \sum_{{\bf R}_k}  \frac{erf(
\sqrt{\lambda} | {\bf r} - {\bf R}_k|)}{| {\bf r} - {\bf R}_k|},
\end{align*}
which is exactly the linear superposition of the electrostatic potentials of all
the gaussian charge distributions of the lattice.

We compare again our view with that of Ref.\cite{ber1} (Eqs.(51)-(53))
 and we note that, as in this case the
charge distributions are infinite, they overlap in all the space. That is why
the contribution of the mean square radii can be interpreted in this case as an
"overlap correction", if we are interested in studying a lattice of point
charge.
Note that, if the charge density is finite, the electrostatic potential of all 
the mean square radii is different from zero only inside the source, while for
infinite charge distributions (but rapidly deacreasing at infinity),
 the potential of the mean square radii is significantly different from zero
 only in the region where the charge distribution is significantly different
 from zero. We have obtained in the above equation that, for a charge
 distribution of the form $\rho({\bf r})= C e^{- \lambda r^2}$, the potential of
 its mean square radii is proportional to $\frac{erfc( \sqrt{\lambda} r)}{r}$.

The self-energy of a gaussian charge distribution is $W_s= \frac{\pi^{5/2} C^2}{
\lambda^{5/2} 2^{1/2}}$ and the electrostatic energy of the lattice: 
\[
W= \frac{2 \pi^4 C}{\omega \lambda^3} \sum_{\bf g} \frac{1}{g^2} e^{ -
\frac{g^2}{2 \lambda}} - \frac{\pi^{5/2} C^2}{
\lambda^{5/2} 2^{1/2}}.
\]

\subsubsection{Exponential charge distribution}
We consider now an exponential charge density of the form: $\rho( {\bf r} )
= C e^{- \lambda r}$, where C and $\lambda >0$ are adjustable parameters. A charge
mean square radius of order $n$ has the expression:
\begin{equation}
\overline{r_0^{2n}}= \frac{4\pi C}{\lambda^{2n+3}} \Gamma(2n+3).
\end{equation}
Eq. (\ref{ch}) becomes in this case:
\begin{equation}
e^{-\lambda r} = 8 \pi \sum_{n=0}^{\infty} \frac{n+1}{ \lambda^{2n+3}} \Delta^n
\delta( {\bf r} )
\end{equation}
and, after taking the Fourier transform, we find the usual Taylor expansion:
\begin{equation}
\frac{1}{ \left( 1+ \frac{k^2}{\lambda^2} \right)^2} = \sum_{n=0}^{\infty}
(-1)^n (n+1) \left( \frac{k^2}{\lambda^2}  \right)^n.
\end{equation}
The electrostatic potential produced by such a lattice at an arbitrary point
${\bf r}$ is:
\begin{align}
\phi( {\bf r} ) = &\frac{8\pi C}{\lambda^3} \sum_{ {\bf R}_k} \frac{1}{ | {\bf r} -
{\bf R}_k |} + \frac{ 32\pi^2 C}{\Omega \lambda^3} \cdot \nonumber \\
&\sum_{ {\bf g}}
\frac{1}{g^2} \left[ \frac{\lambda^4}{(\lambda^2 + g^2)^2}-1\right] 
e^{i {\bf g}{\bf r}},
\end{align}
or, as a single sum in the reciprocal space,
\begin{equation}
\phi( {\bf r} ) = \frac{32 \pi^2 C}{\lambda^3 \Omega}
 \sum_{ {\bf g}} \frac{1}{g^2}
  \frac{\lambda^4}{(\lambda^2 + g^2)^2} e^{i {\bf g}{\bf r}}.
\end{equation}
The self-energy of such a charge distribution is $ W_s = \frac{10 \pi^2
C^2}{\lambda^5} $ and the electrostatic energy of the lattice  is:
\begin{equation}
W= \frac{128 \pi^3 \lambda^2 C^2}{\Omega} \sum_{{\bf g}} \frac{1}{g^2
(\lambda^2+ g^2)^4} - \frac{10 \pi^2 C^2}{\lambda^5}.
\end{equation}

\subsubsection{Slater-type charge distribution without angular dependence}
In Ref. \cite{sla} Slater proposed a radial wave function appropriate for
many-electron atoms and ions of the form:
\[ \psi(r) = r^{n^*-1} e^{-\frac{Z-s}{n^*}r},\] where $n^*$ is an effective
quantum number, $Z$ is the atomic number
 and $s$ a screening constant. We consider now that at each node
of our Bravais lattice there is a charge density which corresponds to Slater's
radial wave function:
\begin{equation}
\rho(r)= C r^{2(n^*-1)} e^{-2 \lambda r},
\end{equation}
where $\lambda \equiv \frac{Z-s}{n^*}  $ and $C$ is a constant.
The mean square charge radii of order $p$ which corresponds to such a charge
density is: \[ \overline{r_0^{2p}}= \frac{4\pi C(2p+2n^*)!}{(2\lambda)^{2p+2n^*+1}}.\]
Eq.(\ref{ch}) becomes in this case:
\begin{equation}
r^{2(n^*-1)} e^{-2\lambda r} = 4\pi\sum_{p=0}^{\infty} \frac{ (2p+2n^*)!}{ 2^p p!
(2p+1)!! (2\lambda)^{2p+ 2n^* +1}} \Delta^p \delta( {\bf r})
\end{equation}
and, after taking the Fourier transform:
\begin{align*}
\Gamma(2n^*+1) _2F_1 \left( \frac{2n^*+1}{2}, n^*+1; \frac{3}{2}; -
\frac{k^2}{4\lambda^2}  \right) = \\ \sum_{p=0}^{\infty} \frac{ (2p+2n*)! (-1)^p
k^{2p}}{2^p p! (2p+1)!! (2\lambda)^{2p}},
\end{align*}
which is exactly the defining series of the Gauss hypergeometric function.
The electrostatic potential at an arbitrary point of such a lattice is:
\begin{align*}
\phi({\bf r}) =\frac{ 4 \pi C(2n^*)!}{ (2 \lambda)^{2n^*+1}} \sum_{ {\bf R}_k} \frac{1}{ |
{\bf r} - {\bf R}_k| } + 
\frac{16 \pi^2 C (2n^*)!}{\Omega (2 \lambda)^{2n^*+1}} \cdot \\
\sum_{
{\bf g}} \frac{1}{g^2} \Bigg\{ \frac{i \lambda}{2 n^* g} \Big[ 
\Big( \frac{2\lambda}{ 2 \lambda
+ig} \Big)^{2n^*} - \Big( \frac{2\lambda}{ 2 \lambda -ig} \Big)^{2n^*} \Big] -1 \Bigg\} e^{i {\bf g} {\bf
r}},
\end{align*}
or, as a single sum in the reciprocal space:
\begin{align}
\phi({\bf r}) = & \frac{16 \pi^2 C(2n^*)!}{\Omega (2 \lambda)^{2n^*+1}}\cdot \nonumber \\
& \sum_{
{\bf g}}  \frac{i \lambda}{2 n^* g^3} \Big[ \Big( \frac{2\lambda}{ 2 \lambda
+ig} \Big)^{2n^*} - \Big( \frac{2\lambda}{ 2 \lambda -ig} \Big)^{2n^*} \Big]  e^{i {\bf g} {\bf
r}}.
\end{align}
The self-energy of such a charge distribution is:
\begin{equation}
W_s= \frac{\pi^2 C^2 \Gamma(2n^*)}{2^{4n^*-1} \lambda^{4n^*+1}} \left[  4n^*
\Gamma(2n^*) - \frac{2}{\sqrt{\pi}} \Gamma\left(2n^* + \frac{1}{2}\right) \right] 
\end{equation}
and the electrostatic energy of the lattice is:
\begin{align*}
W=& \frac{16 \pi^3 i C^2 \left[ \Gamma(2n*) \right]^2}{\Omega}\cdot \\
& \sum_{{\bf g}} 
\frac{ \left[ (2 \lambda -ig)^{2n^*} - (2 \lambda +ig)^{2n^*} \right]}{ g^4 (4
\lambda^2 + g^2)^{3n^*}} sin \left[ 2n^* arctg \frac{g}{2\lambda}  \right]-\\
&\frac{\pi^2 C^2 \Gamma(2n^*)}{2^{4n^*-1} \lambda^{4n^*+1}} \left[  4n^*
\Gamma(2n^*) - \frac{2}{\sqrt{\pi}} \Gamma\left(2n^* + \frac{1}{2}\right) \right].
\end{align*}

\section{Conclusions}
We have written the electrostatic potential and the electrostatic
interaction energy in a Bravais lattice of arbitrary charge
distributions as  sums in the reciprocal space. We have
shown that the so-called overlapping correction, which have been written
in Ref. \cite{ber1} as a sum in the direct space, is in
fact the contribution of the mean square radii to the 
interaction energy.

We have considered in this paper only charge densities with spherical symmetry.
In the case of charge distributions which have angular dependence do not appear
important technical complications and in many cases the results are finite
linear combinations of the results above obtained, with some weights given by
the values of the angular integrals. For example, if we consider a Slater-type
charge distribution of the type $ \rho( {\bf r}) = C r^{2(n^*-1)} e^{ -2 \lambda
r} | Y_{1m}(\theta, \phi)|^2$, where  $Y_{1m}, \;m=0, \pm 1$ is a spherical
harmonic (\cite{gra}), the only non-zero multipoles are
the monopole and the quadrupole together with their infinite number of mean
square radii. Therefore, after we have found the weights of these two type of
multipoles, the calculation follows as in Sec. 2.3.3 and the final result can be
written as a linear superposition. Such elaborate calculations will be the
subject of an other paper.

\end{document}